\def\year{2021}\relax
\documentclass[letterpaper]{article} 
\usepackage{aaai21}  
\usepackage{times}  
\usepackage{helvet} 
\usepackage{courier}  
\usepackage[hyphens]{url}  
\usepackage{graphicx} 
\usepackage{natbib}  
\usepackage{caption} 
\usepackage{graphicx}
\usepackage{array}
\usepackage[dvipsnames,usenames]{color}
\newcolumntype{P}[1]{>{\raggedright\arraybackslash}p{#1}}

\title{Using Transformers to Provide Teachers with Personalized Feedback on their Classroom Discourse: The TalkMoves Application}
\author{
    Abhijit Suresh,  
    Jennifer Jacobs,
    Vivian Lai,
    Chenhao Tan, \\
    Wayne Ward,
    James H. Martin, Tamara Sumner
    \\
   
}
\affiliations{

    \textsuperscript{\rm 1} Institute of Cognitive Science\\
    Department of Computer Science \\
    University of Colorado Boulder\\
    FirstName.LastName@colorado.edu

}

\begin{document}

\maketitle
\begin{abstract}
TalkMoves is an innovative application designed to support K-12 mathematics teachers to reflect on, and continuously improve their instructional practices. This application combines state-of-the-art natural language processing capabilities with automated speech recognition to automatically analyze classroom recordings and provide teachers with personalized feedback on their use of specific types of discourse aimed at broadening and deepening classroom conversations about mathematics. These specific discourse strategies are referred to as “talk moves” within the mathematics education community and prior research has documented the ways in which systematic use of these discourse strategies can positively impact student engagement and learning. In this article, we describe the TalkMoves application’s cloud-based infrastructure for managing and processing classroom recordings, and its interface for providing teachers with feedback on their use of talk moves during individual teaching episodes. We present the series of model architectures we developed, and the studies we conducted, to develop our best-performing, transformer-based model (F1 = 79.3\%). We also discuss several technical challenges that need to be addressed when working with real-world speech and language data from noisy K-12 classrooms.
\end{abstract}

\section{Introduction}
\label{intro}
The TalkMoves application builds on advances in deep learning for natural language processing and speech recognition to automatically analyze classroom recordings and provide K-12 teachers with personalized feedback on their instructional practices. Classroom recordings consist of video, audio, and/or transcripts of teaching episodes, including entire lessons or portions of lessons. In this research, we provide teachers with off-the-shelf tools such as tablets and SWIVL devices \cite{franklin2018using,mccoy2018case} that enable them to self-record high-quality video and audio in noisy classroom environments. Much of the critical information from these classroom recordings of teacher and student interactions is captured in the speech and language components. The TalkMoves application processes each classroom recording by analyzing every teacher and student utterance in order to generate a detailed record of the “talk moves” being used in classroom conversations along with other relevant discursive features. The application then provides teachers with detailed feedback on the degree to which they engaged their students in productive patterns of discourse.

The purpose of the TalkMoves application is to address a significant challenge in mathematics education: providing teachers with immediate and actionable feedback on their use of effective classroom discourse strategies. Currently, providing teachers with such feedback requires highly trained observers to hand code transcripts of classroom recordings using qualitative research methods (e.g., \cite{correnti2015improving}). This approach is time-consuming, expensive, and demands considerable human expertise. As a result, current approaches simply do not scale to large numbers of teachers. TalkBack will automate and scale up this process, enabling more teachers to receive prompt and accessible feedback on these important instructional practices.

Notably, from a natural language processing perspective, mathematics education research has converged on a detailed understanding of the types of discourse strategies that promote student learning and engagement, and several groups have developed detailed frameworks describing these strategies and how to best use them \cite{zhang2004can,szymanski2002producing}.Talk moves are specific discussion strategies that teachers can use to enable all students to equitably participate in a rigorous classroom learning environment. Teachers use talk moves to encourage their students to contribute and listen to each other, to engage with the math content, and to dig deeply into their own reasoning. In the studies presented here, we are building on a well-established and well specified talk moves framework known as Accountable Talk \cite{o2015scaling}. Accountable talk looks “strikingly similar to the norms of discourse called for in theories of deliberative democracy”  \cite{michaels2008deliberative}. Specifically, accountable talk supports a discussion-based classroom community with the expectation that all students have equal access to participation, subject matter content, and developing appropriate habits of mind \cite{michaels2010accountable}.

In our previous work, we trained a deep learning model based on Bidirectional Long Short-Term memory (Bi-LSTM) to label all the teacher sentences spoken during math lessons with their corresponding Accountable Talk move and achieved an F1 performance up to 65\% \cite{suresh2018using, suresh2019automating}. The noisy and imbalanced nature of classroom speech data can be challenging when performing downstream sequence classification. We have leveraged recent advances in natural language processing, including contextual word embedding \cite{pennington2014glove} and transformers \cite{devlin2018bert, liu2019roberta}, to develop and study a series of model architectures to classify student-teacher sequences containing Accountable Talk moves. Results show a significant improvement over our previous work, with an F1 performance of 79.3\%. We discuss several technical challenges arising from working with speech and language data collected in real-world classrooms, such as widely varying use of different talk move types and the impact of automated speech recognition on talk move model classification errors.

%
%
    %
    %
    %
    %
    %
    %

\section{Related Educational Theory}

The Common Core State Standards (CCSS) for mathematics underscore the need for social interaction and communication as a means to promote learning environments in which students actively contribute and engage with  each other’s ideas  \cite{franke2015student}. Michaels, O’Connor and colleagues developed an approach to classroom discourse called “accountable talk”  \cite{o2015scaling}. At the heart of accountable talk theory is the notion that teachers should organize discussions that promote students’ equitable participation in a rigorous learning environment. The use of talk moves is an “important and universally recognized dimension of teaching” \cite{correnti2015improving}, and prior research has established strong linkages between productive classroom discourse and student achievement  e.g., \cite{boston2012assessing, munter2014developing, resnick2010well, walshaw2008teacher, webb2014engaging}. 

Intentionally and skillfully using talk moves takes time and practice \cite{correnti2015improving}.  However, using talk moves helps to ensure that classroom discussions will be purposeful, coherent, and productive. As shown in Table 1, talk moves can be classified as falling into three broad categories: accountability to the learning community, accountability to content knowledge, and accountability to rigorous thinking. The goal is for teachers to utilize a variety of talk moves, as appropriate, within each of these categories to ensure that students are engaged and actively participating, responsible for making accurate and appropriate claims, and providing accessible and well-reasoned arguments \cite{michaels2010accountable}.

\begin{table*}[h]
\begin{center}
\begin{tabular}{|P{3cm}|P{3cm}|p{5cm}|p{5cm}|} 
 \hline
\bf Category & \bf Talk move & \bf Description & \bf Example \\ [1ex] 
\hline
Learning Community & Keeping everyone together &  Prompting students to be active listeners and orienting students to each other & “What did Eliza just say her equation was?”\\ [1ex] 
\hline
Learning Community & Getting students to another’s ideas & Prompting students to react to what a classmate said & “Do you agree with Juan that the answer is 7/10?” \\ [1ex] 
\hline
Learning Community & Restating & Repeating all or part of what a student says word for word & “Add two here" \\ [1ex] 
\hline
Content Knowledge& Press for accuracy & Prompting students to make a mathematical contribution or use mathematical language &“Can you give an example of an ordered pair?” \\ [1ex] 
\hline
Rigorous thinking & Revoicing &Repeating what a student says but adding on to it or changing the wording &“Julia told us she would add two here.” \\ [1ex] 
\hline
Rigorous thinking & Press for reasoning &Prompting students to explain or provide evidence, share their thinking behind a decision, or connect ideas or representations& “Why could I argue that the slope should be increasing?” \\ [1ex] 
\hline
\end{tabular}
\caption{\label{font-table} The six accountable teacher talk moves incorporated in the TalkMoves application.}
\end{center}
\label{table:desc}
\end{table*}

\section{Talk Moves Model}
The primary goal of this study is to classify teacher sentences into six discourse classes or labels with high reliability in order to generate feedback on individual teacher’s instruction. In addition, the model should also be able to distinguish between teacher sentences with and without talk moves. We define these efforts as a 7-way sequence classification problem i.e., for each teacher sentence in the transcript, the model produces a  probability (softmax) distribution over the six discourse strategies and “None”. Our previous attempt \cite{suresh2019automating} to classify teacher sentences relied on a turn-based format, where each turn was defined as a spoken exchange between the teacher and a student. We used multiple features including sentence embedding, bag-of-word embedding with GloVe \cite{pennington2014glove} and a count-vectorizer. The resulting model had a F1 performance accuracy up to 65\%. In an effort to improve the robustness, reliability and performance of the model, we have now extended this work using more updated, state-of-the art models to detect talk moves, such as transformer architectures. Recent advances in transformer architectures and its variants have resulted in significant improvements in performance across a number of downstream tasks including similarity, inference, paraphrasing and classification tasks among others \cite{devlin2018bert, liu2019roberta}. In this section, we discuss the talk moves data, the evaluation metrics, and the results from different model experiments and architectures.

\subsection{Data}
For this study, we collected 501 written transcripts of kindergarten through 12th grade (K-12) math lessons from multiple sources. All the transcripts were segmented into sentences using an automated script. Each sentence in the transcript was manually coded for six talk moves by one of two experienced annotators who were extensively trained on accountable talk and adhered to a detailed coding manual. The annotators established initial reliability with one another prior to applying the codes and again when they were approximately halfway through coding to ensure that their coding remained accurate and consistent. Inter-rater agreement, calculated using Cohen’s kappa  \cite{mchugh2012interrater} , was above .90 for each talk move at both time periods (see Table 2). These sentences annotated by human experts served as the “ground-truth” training dataset for our model.

\begin{table*}[h]
\begin{center}
\begin{tabular}{|p{5cm}| p{2cm}| p{2cm}| p{2cm}|} 
 \hline
\bf Coding decision & \bf Inter-rater agreement & \bf Initial kappa & \bf Midpoint kappa  \\ [0.5ex] 
 \hline
 Keeping everyone together & 88\% & 0.91 & 0.96\\
 \hline
 Getting students to relate & 94\% & 0.91 & 0.92\\
 \hline
  Restating & 100\% & 1.0 & 1.0 \\
 \hline
  Revoicing & 98\%  & 0.99 & 1.0\\
 \hline
  Press for accuracy & 89\% & 0.93 & 0.95\\
 \hline
  Press for reasoning & 92\% & 0.95& 0.95 \\ [1ex] 
 \hline
\end{tabular}
\end{center}
\caption{\label{font-table}  Cohen’s kappa scores between annotators who labelled each sentence from the collected transcripts as one of 7 unique labels (6 talk moves and “none”).}
\end{table*}

All the sentences in the dataset were stripped of punctuation and lower-cased. In this study we used a student-teacher “sentence pair” format, which is a combination of a teacher sentence concatenated with the immediately prior student sentence. This format enables the model to have access to the previous student sentence as context, which is especially important for the talk moves Restating and Revoicing (when the teacher essentially repeats what a student has already said). Examples of student-teacher sentence pairs are shown in Table 3.

\begin{table*}[h]
\begin{tabular}{|p{0.5\textwidth} p{0.5\textwidth}|} 
 \hline
\multicolumn{2}{|c|}{\bf Example of data organized as a turn} \\
\hline
\multicolumn{2}{|l|}{student: so you put the eight on the box}\\
 \hline
\multicolumn{2}{|l|}{student: then you get eight}\\
 \hline
\multicolumn{2}{|l|}{teacher: oh so you were using this side to help you get that side}\\
 \hline
\multicolumn{2}{|l|}{teacher: let me see if i can figure out what you said} \\
 \hline
 \multicolumn{2}{|c|}{\bf Example  of data organized  as sentence pairs} \\
 \hline
  student: then you get eight & teacher: oh so you were using this side to help you get that side \\
  \hline
   student: - & teacher: let me see if i can figure out what you said  \\
   \hline
  student: then another line going straight down & teacher: can you go ahead and explain what you did \\
 \hline
\end{tabular}
\caption{\label{font-table} Example of data organized as turns \cite{suresh2018using} compared to sentence pairs.}
\end{table*}

The dataset used in this study consists of 176,757 sentences, which can be broken down into 115,418 teacher sentences and 61,339 student sentences. The skewed distribution of the individual talk moves makes it harder for the model to differentiate between a high frequency label compared to a low frequency label (see Figure 1). In addition, sentences extracted from classroom transcripts are noisy, meaning they frequently lack well-formed syntax, include misspellings, and have missing words. The unbalanced distribution along with the noisy nature of the individual sentences make talk moves classification a challenging sequence classification problem. The talk moves dataset was split into training, validation, and test sets according to an 80/10/10 \% split. Both the validation and testing sets were stratified to mimic the distribution of the labels in the training set. The validation set was used for hyper-parameter tuning and the testing set was used for evaluating the model performance.

\begin{figure}[h!t]
\center
  \includegraphics[scale=0.9]{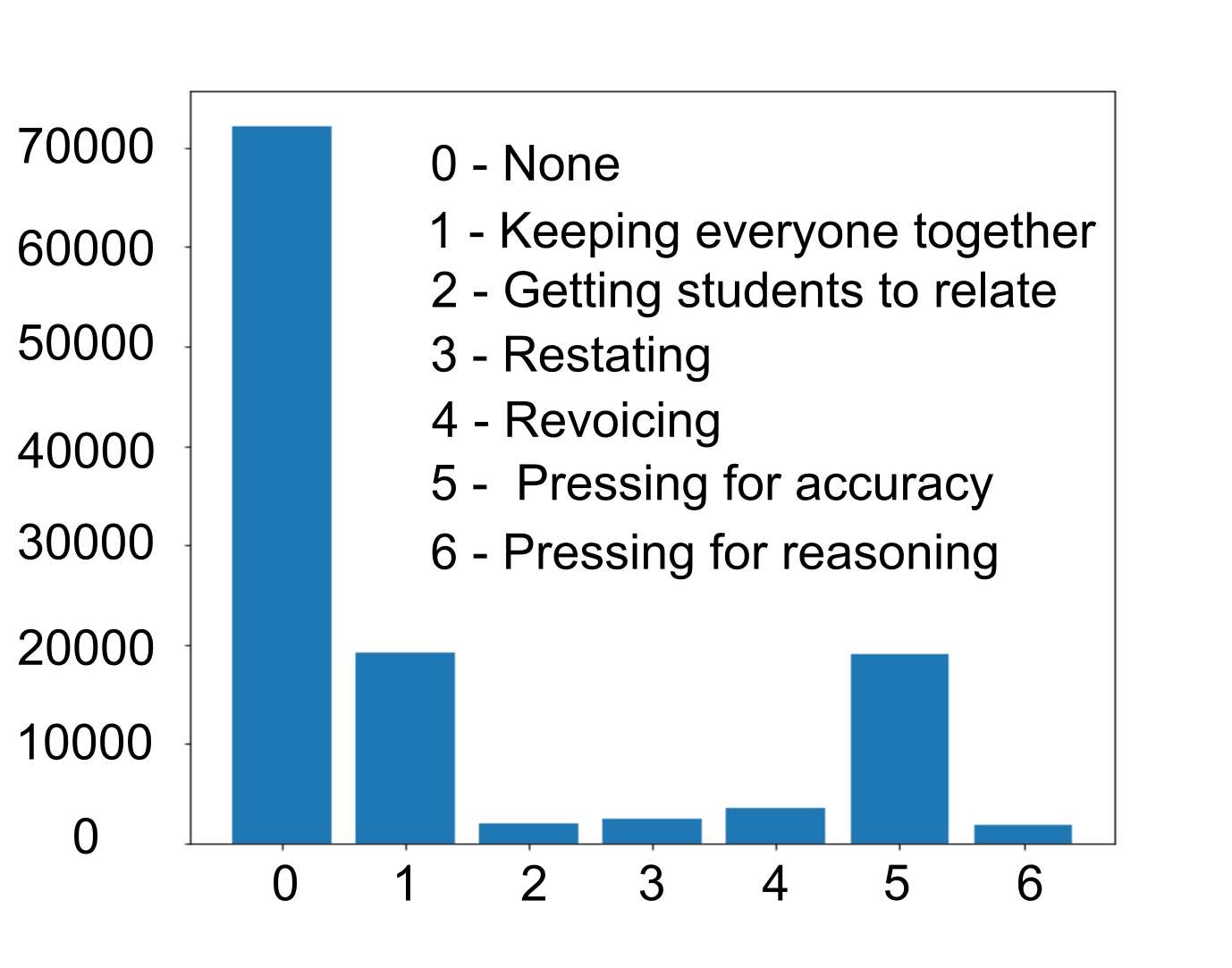}
  \caption{Skewed distribution of the TalkMoves data set}
  \label{}
\end{figure}

\subsection{Metrics}
In order to determine the performance of a given model across different talk moves, we need a reliable statistical measure. Post-training the dataset, we examined each model’s performance on the test set to yield a confusion matrix which allowed us to calculate two different metrics: a Matthew Correlation Coefficient (MCC) and an F1 measure. We opted for these metrics rather than simply calculating the model’s degree of accuracy due to the fact that the data had an unbalanced distribution of labels (see Figure 1). Typically, an F1 score is  calculated as the harmonic mean of precision and recall. The reported F1 score in our study was calculated across 6 labels (ignoring the “None” label) as an indicator of model performance across the talk moves. Similar to F1, MCC has a range between 1 to -1, with 1 indicating a perfect classifier and -1 refering to the wrong classifier. In recent studies, MCC scores have been proven to be more stable and reliable than F1 \cite{chicco2020advantages} and to better reflect the performance of a model. Although MCC was originally designed for binary classifiers, it can be extended to multi-class scenarios such as in our study. In this paper we present both MCC and F1 scores, which for our model experiments are generally in agreement. 

\subsection{Model Experiments}
The goal of the talk moves classifier is to predict the label associated with each student-teacher sentence pair. The predicted labels can then be used to generate automated feedback for teachers on their classroom discourse. We began with a Bi-LSTM network with GloVe embeddings \cite{pennington2014glove} to represent all the sentences in the embedding space. LSTMs, in general, were designed to perform better than recurrent neural networks (RNNs) in capturing long term dependencies \cite{sherstinsky2020fundamentals}. This model produced a F1 score of 72.26\% as seen in Table 4. All the reported scores reflect model performance on the test set.

Following the BiLSTM models, we experimented with attention mechanisms, which originate from the domain of neural machine translation \cite{chu2010new}. Adding an attention layer on top of the Bi-LSTM enables the neural network model to focus on specific input words relative to others in the sentence. The resulting model showed only a marginal improvement in performance. Additionally, we explored transformers. Leveraging the encoder block from the transformer architecture \cite{vaswani2017attention}, Devlin and colleagues \cite{devlin2018bert} introduced Bidirectional Encoder Representations from Transformers or BERT, a language-based model pre-trained on unlabeled data. Pre-trained BERT can be finetuned with the addition of an output layer to create state-of the art models applied to downstream tasks like sequence classification, similarity analysis and inference tasks \cite{wang2018glue}. The advent of BERT revolutionized the field of natural language processing and led to the introduction of variants such as XLNET \cite{yang2019xlnet}, Roberta \cite{liu2019roberta}, and Albert \cite{lan2019albert}. Differences between these variants include the data used for pre-training, different ways of masking parts of the input, and hyperparameters such as maximum sequence length. In this study, we began with fine-tuning BERT followed by its variants on the TalkMoves data.

\section{Parameter Selection}

Hyper-parameter tuning is an important step to identify the best version of a model within the context of the dataset. Some of the models such as BERT-LARGE, ROBERTA-LARGE and ALBERT-BASE are very sensitive to different parameters. We considered the following variables for parameter tuning: learning rate (2e-5, 3e-5, 4e-5, 5, e-5), number of epochs (3-6), batch size (4, 8, 16, 32), warmup steps (0, 100, 1000) and maximum sequence length (128, 256, 512). We trained the model multiple times with an exhaustive choice of these parameters using Amazon EC2 instance (p3dn.24xlarge) with 8 Tesla V100 GPUs in parallel. We also used mixed precision (fp16) to speed up the training process 
\cite{haidar2018harnessing}. The code was implemented in Python 3.7 with Pytorch. ROBERTA-LARGE had the best performance on the test set. However, to optimize computation time, a fine-tuned DISTILROBERTA-BASE was incorporated into the TalkMoves application pipeline.

\begin{table}[h]
\begin{center}
\begin{tabular}{|p{5cm}| p{1cm} |p{1cm}|} 
\hline
\bf Model & \bf F1 \% & \bf MCC \\
\hline
BASE-MODEL \cite{suresh2019automating} & 65 &   - \\
\hline
BiLSTM with GloVe embeddings & 72.26 &  0.7042 \\
\hline
BiLSTM with Attention and GloVe embeddings & 72.64 &  0.7072 \\
\hline
BERT-BASE \cite{devlin2018bert} & 78.89 &   0.7718 \\
\hline
ALBERT-BASE \cite{lan2019albert} & 78.18 & 0.7637 \\
\hline
ROBERTA-BASE \cite{liu2019roberta} & 78.94 & 0.7704 \\
\hline
XLM-ROBERTA-BASE \cite{conneau2019unsupervised} & 78.66 & 0.7684 \\
\hline
BERT-LARGE \cite{devlin2018bert} & 79.04 & 0.7774 \\
\hline
ROBERTA-LARGE \cite{liu2019roberta} & 79.33 &  0.7779 \\
\hline
XLNET-BASE \cite{yang2019xlnet} & 78.29 & 0.7672 \\
\hline
DISTILBERT-BASE \cite{sanh2019distilbert} &  78.02 & 0.7616 \\
\hline
DISTILROBERTA-BASE \cite{sanh2019distilbert}  & 77.90 & 0.7641 \\
\hline
\end{tabular}
\end{center}
\caption{\label{font-table} Results from different model experiments and their corresponding F1 performance and MCC scores on TalkMoves test set. ROBERTA-LARGE had the best performance on the test set. To optimize computation time, fine-tuned DISTILROBERTA-BASE is incorporated into the TalkMoves application}
\end{table}

\subsection{Designing the Automated Feedback}

\begin{figure*}[h!t]
\center
  \includegraphics{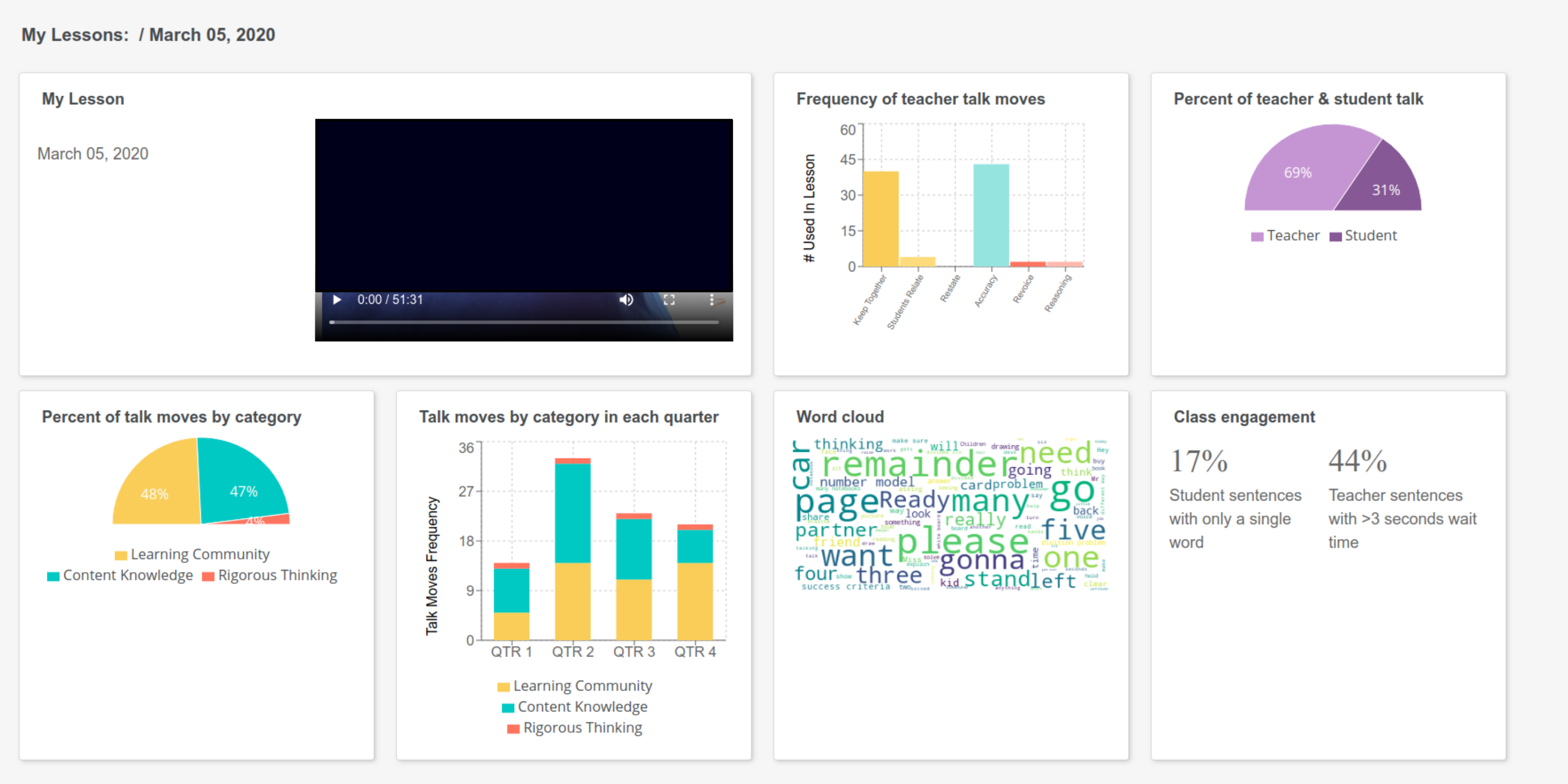}
  \caption{A screenshot of the front-end feedback interface for teachers for a single classroom recording}
  \label{}
\end{figure*}

The final step in the development of the TalkMoves application was to specify the nature of the feedback that teachers would receive. This feedback primarily relates to the six talk moves, but also includes other information about the classroom discourse such as the amount of teacher versus student talk, the number of one-word student utterances, and the frequency of teachers’ sentences with wait time. In addition, the application provides feedback on teachers’ individual lessons along with changes in their lessons over time. The project convened a teacher advisory board to collaboratively
brainstorm suggestions and capture teachers’ reactions to initial visualizations of the feedback and mock-ups of the application design. Based on the ideas generated by the advisory board, the project team designed an interactive “dashboard” for each lesson to display selected analytics using a variety of graphics and visual representations (see Figure 2). 

 In the current version of the application, for each uploaded lesson the dashboard displays (1) video of the lesson, (2) the frequency of each talk move and the total number of talk moves, (3) the percentage of teacher and student talk, (4) the percentage of talk moves within each of three categories, (5) the amount of talk moves by category during each quarter of the lesson, (6) a word cloud showing the most frequently used words, and (7) the percentages of students’ one word responses and teacher sentences with at least 3 seconds of wait time(to allow for student contributions). The interface also includes a “teacher guide” that contains information about accountable talk theory, definitions and examples of each talk move, and how the application was developed. 
 
 This first version of the TalkMoves application includes a subset of the intended features and pages, and at present is only available to a small group of pilot teachers in two midwestern school districts in the United States of America. This group of 21 teachers, serving grades 3-12, used the application throughout the 2019-2020 academic year. Each teacher recorded between 3-15 lessons, viewed their feedback, completed surveys and participated in interviews with members of the research team. Based on the teachers’ insights and concerns, a second version of the application is currently underway.
\section{System architecture and Implementation}

\begin{figure*}[t]
\center
  \includegraphics[scale=.45]{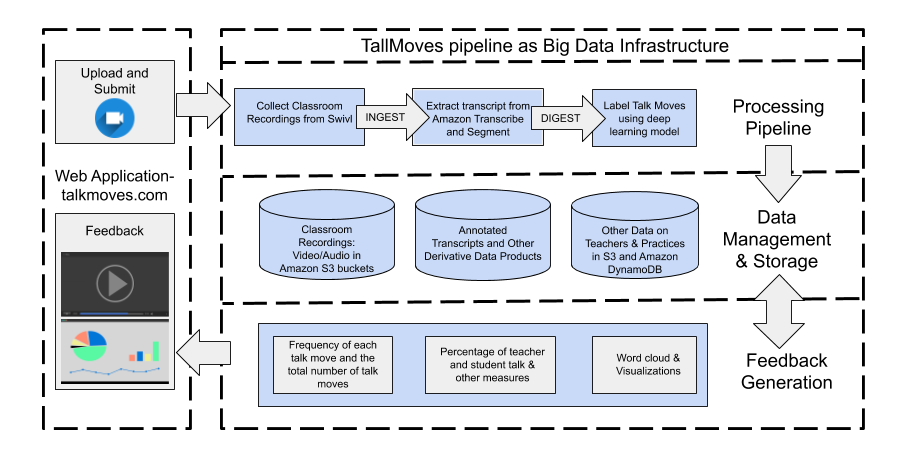}
  \caption{TalkMoves System architecture. All the modules are processed using Amazon web services (AWS)}
  \label{}
\end{figure*}

The TalkMoves application infrastructure has been designed to asynchronously process classroom recordings to generate personalized feedback using Amazon Web Services (AWS). Classroom video and audio recordings are collected using a hardware device called Swivl, designed to provide automated video capture for K-12 classrooms \cite{franklin2018using}. Each teacher participating in the Talk Moves project is equipped with an iPad and a Swivl. The Swivl is used to capture the classroom session through a mounted iPad that swivels with the teacher as they move around the classroom. Teachers were also given five audio recording markers; one was meant to be worn around the teacher’s neck and four were to be distributed around the classroom or near students’ desks. At the start of the class, the teacher can begin recording using the Swivl application on their iPad. Once they are finished recording, the teacher can rename the video file and it  will be automatically uploaded to the Swivl cloud. The TalkMoves system then collects the data from the Swivl cloud, processing one video at a time through the Talkmoves pipeline. The system architecture of the TalkMoves pipeline is summarized in Figure 3. The audio from classroom recordings is converted into written transcripts using Amazon Transcribe, which are then processed with a deep learning model to identify whether there is a talk move corresponding to each teacher sentence. The system then generates feedback based on the output from the model, which is presented to the teachers using a web interface.

\section{Discussion}
The TalkMoves application was designed to provide teachers with feedback that can promote rich discussions in their mathematics classrooms. By combining contemporary tools from data science and natural language processing, we have created a system which can provide feedback that is generally on par with domain based instructional experts in terms of reliability. In mathematics education and teacher classroom observation literature, inter-rater agreement of approximately 80\% is the generally agreed-upon threshold \cite{wilhelm2018exploring}, although lower scores are certainly possible even among well-trained raters \cite{hill2012rater}. In this context, model performance of 79.33\% highlights the potential of integrating natural language processing within an educational innovation.

Building on our initial turn-based BiLSTM model which had an FI score of 65\%, we have made significant progress towards creating a robust model that better approximates expert human behavior. To take a closer look at how well the new model compares to human raters, we performed a detailed error analysis. Error analysis includes analyzing and identifying patterns in example sentences  that are misclassified for each talk move. Among the six talk moves, “Keeping everyone together”, “Getting students to relate” and “Revoicing” have the lowest individual F1 scores of 75\%, 73\% and 69\% respectively. We conjecture that the accuracy of these individual talk moves, as well as the overall performance of the system, may be improved by increasing the context window available for classifying the teacher sentences as opposed to the present setup where each teacher sentence is preceded by a single student sentence.

One important limitation of this work is the challenge of ASR systems to accurately recognize young children’s speech. In particular we have found that student talk is severely underestimated by Amazon Transcribe, likely due to low confidence levels and errors brought on by acoustic variability, unpredictable articulation, and other behaviors common in children’s language production \cite{booth2020evaluating, gerosa2007acoustic}.

\section{Conclusion}
This study contributes to an increasing body of literature on the development of automated tools that have strong potential to support teachers’ professional learning  \cite{killion2012meet}. Other research teams have successfully used a combination of speech processing and supervised machine learning to discriminate basic classroom discourse structures such as lecture and group work \cite{donnelly2016automatic, owens2017classroom, wang2014automatic} and to predict the occurrence of discursive practices such as the teacher’s instructional talk and questions \cite{owens2017classroom, jensen2020toward}.The work presented in this paper extends these efforts in several ways by incorporating new approaches to use AI tools for K-12 education that serve (1) as a domain expert providing automated feedback on classroom instruction, (2) as an application of the latest NLP models applied to interpreting complex, large scale patterns in classroom transcripts, and (3) as an end-to-end system designed to support teachers to lead discourse-rich mathematics lessons.

\section*{Acknowledgements}
The research team would like to thank Eddie Dombower and his team at Curve 10 for their contributions to the design and implementation of the TalkBack application. This material is based upon work supported by the National Science Foundation under Grant No.1837986 : The TalkBack Application: Automating Analysis and Feedback to Improve Mathematics Teachers’ Classroom Discourse.

\bibliographystyle{aaai21}
\bibliography{aaai21}

\end{document}